\documentclass[aps,twocolumn,showpacs,floatfix,superscriptaddress]{revtex4}
\usepackage{graphicx}
\usepackage{amsmath}
\usepackage{amsfonts}
\usepackage{amssymb}
\usepackage{epsfig}

\begin{document}

\newcommand{\la}{\langle}
\newcommand{\ra}{\rangle}
\def\be{\begin{equation}}
\def\ee{\end{equation}}
\def\bea{\begin{eqnarray}}
\def\eea{\end{eqnarray}}
\def\bma{\begin{mathletters}}
\def\ema{\end{mathletters}}
\newcommand{\one}{\mbox{$1 \hspace{-1.0mm}  {\bf l}$}}
\newcommand{\eins}{\mbox{$1 \hspace{-1.0mm}  {\bf l}$}}
\def\C{\hbox{$\mit I$\kern-.7em$\mit C$}}
\newcommand{\tr}{{\rm tr}}
\newcommand{\half}{\mbox{$\textstyle \frac{1}{2}$}}
\newcommand{\shalf}{\mbox{$\textstyle \frac{1}{\sqrt{2}}$}}
\newcommand{\ket}[1]{ | \, #1  \rangle}
\newcommand{\bra}[1]{ \langle #1 \,  |}
\newcommand{\proj}[1]{\ket{#1}\bra{#1}}
\newcommand{\kb}[2]{\ket{#1}\bra{#2}}
\newcommand{\bk}[2]{\langle \, #1 | \, #2 \rangle}
\def\II{I(\{p_k\},\{\rho_k\})}
\def\ss{{\cal K}}
\tolerance = 10000




\title{Adiabatic Path to Fractional Quantum Hall States of a Few Bosonic Atoms}
\author{M. Popp}
\affiliation{Max--Planck Institute for Quantum Optics,
Garching, Germany}
\author{B. Paredes}
\affiliation{Max--Planck Institute for Quantum Optics, Garching, Germany}
\author{J. I. Cirac}
\affiliation{Max--Planck Institute for Quantum Optics,
Garching, Germany}

\begin{abstract}
We propose a realistic scheme to create motional entangled states of a few 
bosonic atoms. It can experimentally be realized with a gas of ultra cold
 bosonic atoms trapped in a deep optical lattice potential.  By 
simultaneously deforming and rotating the trapping potential on each lattice 
site it is feasible to adiabatically create a variety of entangled states on 
each lattice well. We  fully address the case of $N=2$ and $N=4$ atoms per 
well and identify a sequence of fractional quantum Hall  states: the Pfaffian 
state, the $1/2$-Laughlin quasiparticle and the $1/2$-Laughlin state. Exact 
knowledge of the spectrum has allowed us to design adiabatic paths to these
 states, with all times and parameters well within the reach of current
 experimental setups. 
 We further discuss the detection  of these  states by measuring
different properties as their density profile,  angular momentum
or correlation functions.

\end{abstract}

\date{\today}

 \maketitle

\section{Introduction}
The creation of highly entangled multiparticle states is one of
the most challenging goals of modern experimental quantum
mechanics. In this respect atomic systems offer a very promising
arena in which entangled states can be created and manipulated
with a high degree of control. The experimental difficulty
increases, however, with the number of particles that are to be
entangled, since the system becomes then more sensitive to
decoherence. Starting with a small number of particles as a first
step, important achievements have been already obtained in the
creation of atomic entangled states. For example, in recent
experiments with trapped ions, entangled states of up to four ions
have been demonstrated \cite{Sackett00}. Moreover, in  experiments
with neutral bosonic atoms in optical lattices Bell-type states
have been created by accurately controlling the interactions
between neighbouring atoms \cite{Bloch03}. As a typical feature of
most of the experimentally realized entangled states, atoms get
entangled through their internal degrees of freedom, keeping
separable their motional part.

In this article we develop a scheme to create {\em motional}
entangled states of a small number of atoms in an actual
experimental setup with an optical lattice \cite{Bloch02, Ess04, Ph03}. These
states are a sequence of fractional quantum Hall (FQH) states,
analogous to the ones that appear in the context of fractional
quantum Hall effect \cite{QHEbook}. In contrast to typical atomic
entangled states, the particles are here entangled in real space,
and not in internal space. This peculiarity makes them specially
interesting, for it represents a novel nature of atomic
entanglement.

The possibility of creating FQH atomic states as the Laughlin
state by rapidly rotating the trap confining the atoms has been
discussed in several theoretical works \cite{WG00, P01}. However,
experiments dealing with typically large number of particles have
not yet succeeded in reaching these states. Here, we fully address
the case of a small number  of particles and design a
realistic way of entangling them into FQH states. The experimental
setup that we have in mind corresponds to a situation in which a
Bose-Einstein condensate is loaded in a deep optical lattice. When
the lattice depth is very large tunneling between different sites
is strongly suppressed and the system can be treated as a lattice
of independent wells, each of them with a small number of
particles. By independently rotating each of these 3D wells
\cite{Bpriv} the lowest Landau level (LLL) regime can be achieved
for each copy. We have studied the problem exactly within the LLL
for $N=2$ and $N=4$ particles per well. We have identified a
sequence of highly entangled stable ground states, which are the
Pfaffian state \cite{Pfaffian91}, the $1/2$-Laughlin quasiparticle \cite{L83} and the
$1/2$-Laughlin state \cite{L83}. 
The $1/2$-Laughlin quasiparticle state (which had never been identified before in an atomic system) is particularly interesting. It is the counterpart of the  $1/2$-Laughlin quasihole found in \cite{P01} and contains a  $1/2$-anyon.
Driving the system into these
strongly correlated states is, however,  not trivial. By simply
increasing the frequency of rotation the system will stay in a
trivial non-entangled state with angular momentum zero. Exact
knowledge of the spectrum of the system has allowed us to design
adiabiatic paths to these states by simultaneously rotating and
deforming each of the wells. All parameteres and evolution times
lie well within the reach of present experimental setups. We
further discuss how to detect these entangled states by measuring
different properties as their density profile,  angular momentum
or correlation functions. In particular, we propose a novel
technique to measure the density-density correlation function of
these strongly correlated states. Even though the number of atoms per well is small,  the lattice setup allows to have multiple copies of the system, so that the experimental signal is highly enhanced.

We point out that our findings also show that adiabatically
achieving FQH states for rapidly rotating traps with a large
number of particles turns out to be very challenging, since the relevant experimental parameters scale linearly with the number of particles.
Nevertheless, we hope that our results can shed some light on the
problems that these current experiments are dealing with, and even
may pave the way to new methods of achieving FQH multiparticle
entangled states.

\section{Identification of entangled states}
We consider a system of bosonic atoms loaded in a 3D optical lattice. We assume a commensurate
filling of $N$ atoms per lattice site \cite{Com}, and  a large value of the lattice depth $V_0/E_R \gg
1$, where $E_R=\hbar^2k^2/2M$ is the recoil energy, $k$ is the wave vector of the laser
lattice light, and $M$ the atomic mass. In this limit  the lattice can by treated as  a system of
independent 3D  harmonic wells, each of them having $N$ atoms and a trapping
frequency $\omega \approx \sqrt{V_0E_R}$.

 Let us rotate each of these 3D harmonic
wells around the direction $x_3$ with frequency $\Omega$. We will
identify a sequence of motional entangled ground states of the
$N$ atoms that appear as the frequency $\Omega$ is increased. We
will assume the limit of rapid rotation \cite{P01}. In this case
the motion in the $x_3$ direction is frozen, and the motion in the
plane of rotation $x_1,x_2$ is restricted to the LLL. Note that in
order to project the system onto the LLL we do not need to start
with a 2D configuration (as it is the case in previous proposals
\cite{WG00}), since the fast rotation itself restricts the motion
in the direction of the rotation to zero point oscillations. The
system is then governed by a two dimensional effective
Hamiltonian, which written in units of $\hbar \omega$ has the
form:
\begin{equation}
\label{ham}
H=\left(1-\Omega/\omega\right)L+ 2 \pi \ \eta  V,
\end{equation}
where $L=\sum_{m=0}m \, a^{\dagger}_{m}a_{m}$ is the angular
momentum operator in the $x_3$ direction, and
$V=\sum_{m_1,m_2,m_3,m_4}
V_{m_1,m_2}^{m_3,m_4}a^{\dagger}_{m_1}a^{\dagger}_{m_2}a_{m_3}a_{m_4}$
is the interaction operator. Here the bosonic operator $a^{\dagger}_{m}
(a_{m})$ create (anihilate)
an atom in the state $\vert m \rangle $
of the LLL with well defined $x_3$ component of the angular momentum $m$. The wave functions of the LLL in complex coordinates read
\be
\varphi_m(z)=\langle z \vert m
\rangle = \frac{1}{\sqrt{\pi m!} \ell} \  z^m e^{-\vert z \vert ^2/2} \ ,
\ee
 where
$z=(x_1+ix_2)/\ell$, $\ell=\sqrt{\hbar/M\omega}$, and
$m=0,1,\ldots \infty$. Assuming contact interactions between the
atoms the interaction coefficients are:
\be
V_{m_1,m_2}^{m_3,m_4}=\frac{(m_1+m_2)!}{2^{m_1+m_2}\sqrt
{m_1!m_2!m_3!m_4!}} \ .
\ee
In Hamiltonian (\ref{ham}) we have introduced the 
important interaction  parameter $\eta=\sqrt{2/ \pi} a_s/
\ell$, with $a_s$ the 3D scattering length. Analytical calculations for
scattering potentials of finite size $a_0$ have confirmed that the
pseudo-potential approximation is also valid for tight traps with $a_s \ll \ell$ as long as $a_0 \ll \ell$ is fulfilled \cite{PPCunp}.

\subsection{ N=2}
First we consider the case of two particles per lattice well, which can be
solved analytically. The Hamiltonian (\ref{ham}) is
diagonal in the states $\vert m_r,m_{cm} \rangle$ of well defined
relative ($m_r$) and center of mass ($m_{cm}$) angular momentum:
\begin{equation} \label{H2}
H=\sum_{m_r,m_{cm}}E_{m_r,m_{cm}}\vert m_r,m_{cm}\rangle \langle m_r, m_{cm} \vert,
\end{equation}
with $E_{m_r,m_{cm}}=\delta_ {m_r,0}
\,\eta+(1-\Omega/\omega)(m_r+m_{cm})$. We note that due to the
restriction to s-wave scattering, only particles with zero
relative angular momentum feel the interaction energy. It follows
that for $\Omega/\omega <1-\eta/2$ the ground state of the system
is $\vert 0,0\rangle$ (with total angular momentum $L=0$), which
is not entangled, whereas for $\Omega/\omega >1-\eta/2$ the state
$\vert 2,0\rangle$ (with $L=2$) becomes energetically favourable.
This state, $\langle z_1, z_2 \vert 2,0\rangle \propto
\left(z_1-z_2\right)^2 e^{-\vert z_1 \vert ^2/2} e^{-\vert z_2
\vert ^2/2}$, is clearly entangled since it cannot be written as a
product of two single particle wave functions. It is the Laughlin
state $|\psi_L\rangle$ for two particles at filling factor
$\nu=1/2$ \cite{L83}. In order to quantify the entanglement of
this state we write it in the basis of states $\vert m_1 m_2 \rangle$ with well defined single-particle angular momentum. Then  the Laughlin state takes
the form of a pure two qutrit state: $|\psi_L\rangle= \frac{1}{2}
\left( \vert 02 \rangle + \vert 20 \rangle \right)
-\frac{1}{\sqrt{2}} |11\rangle$. This  is already the  Schmidt decomposition of the state, and the  entropy of entanglement \cite{Bennett96} can immediately calculated to be  $E(|\psi_L \ra)=1.5$. This value is close to $\log_2 3$, corresponding to a maximally entangled pure two qutrit state.  
\subsection{\em N=3}
The case of three particles per lattice well is very similiar to the situation for $N=2$. The $1/2$-Laughlin state ($L=6$) emerges as ground state after an intermediate state with odd angular momentum $L=3$. As we will explain in the next section, ground states with odd angular momentum cannot be reached using our proposal. Hence we now focus on  a setup with four particles per lattice well, for which an interesting sequence of prominent FQH states arises.
\subsection{\em N=4}
\begin{figure}[h]
\centering
\epsfig{file=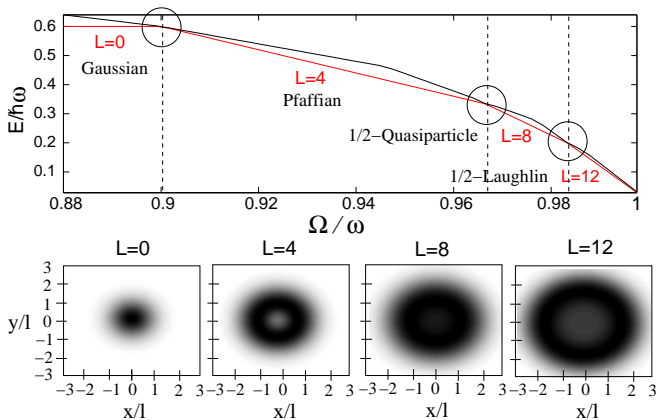,width=\linewidth}
\caption{Lowest two eigenenergies (in units of $\hbar \omega$) of the Hamiltonian (\ref{ham}) for 4 particles and $\eta=0.1$ as  a function of
the trap rotation frequency $\Omega / \omega$.
 The circles mark the
level crossings and $L$ denotes the total angular momentum of the
ground state. The ground state sequence can be
identified as follows (with fidelity given in brackets): L= 0
Gaussian ground state (exact) , L=4 Pfaffian state ($0.95$), L=8
quasiparticle state ($0.98$), L=12 Laughlin state (exact).The
change of angular momentum can readily be obtained from the
increasing width of the  density distribution depicted below.}
\label{Espec4}
\end{figure}

In order to obtain the multi-particle energy spectrum, we have exactly diagonalized the Hamiltonian (\ref{ham}) numerically.
As the frequency of rotation $\Omega$ increases the ground state of the system passes through a sequence of states
with increasing and well defined total angular momentum $L=0, 4, 8, 12$ (see Fig. \ref{Espec4}).
These states can be identified as follows:
The state with $L=0$ is a trivial non-entangled state in which all the atoms are condensed in the
single particle Gaussian state with angular momentum $m=0$.
The first nontrivial ground
state is the $L=4$ state. This state is not, as one might expect, a
single vortex state, in which all the particles would be condensed in the single particle state $m=1$.
In contrast, this state is highly entangled and is very close (fidelity $0.95$) to the
well-known Pfaffian state:
\be
\psi_{Pf}({[z]})= \prod_{i<j}^4 (z_i-z_j) \ \textrm{Pf}\left(
\frac{1}{z_i-z_j}\right) \ .
\ee
 This state is specially interesting,
also in the context of quantum information theory, because its
elementary excitations are known to exhibit non-abelian
statistics \cite{Kit97}. The next stable state in row ($L=8$) can be very well
characterized (fidelity $0.98$) by a Laughlin quasiparticle state: 
\be
\psi_{QP}([z])=\frac{\partial}{\partial z_1}
\ldots  \frac{\partial}{\partial z_4} \ \psi_L \ .
\ee
 This state is
the counterpart of the quasihole excitation, which has previously
been studied in the context of $1/2$-anyons in rotating
Bose-Einstein condensates \cite{P01}. Finally, the last stable
state is identical to the $\frac{1}{2}$-Laughlin state, which we
have already encountered in the case of two particles per well:
\be \label{psiL}
\psi_L([z])= \prod_{i<j}^4 (z_i-z_j)^2  \prod_k^4 e^{|z_k|^2/2} \ .
\ee 
This state is an exact eigenstate of (\ref{ham}) with zero interaction energy. In
Fig. \ref{Espec4} we have plotted the density distribution in the $x_1,x_2$ plane of
the different stable ground states. As the frequency of rotation $\Omega/\omega$ increases
the wave function spreads, and the interaction between the atoms decreases.

\section{Adiabatic paths to entangled states}
 The sequence of
entangled states we have described above cannot be obtained by
simply adiabatically increasing the frequency of rotation
$\Omega$. For the rotational symmetry leads to level crossings
between different angular momentum states (Fig. \ref{Espec4}). In
order to pass adiabatically from the zero angular momentum ground
state to higher angular momentum  states the spherical symmetry of
the trapping potential has to be broken. For our optical lattice
setup this can  be achieved for example by deforming the formerly
isotropic trapping potential on each well and letting the
deformation rotate with frequency $\Omega$ \cite{Bpriv}. In the rotating frame
the new trapping potential has the form $V_p \propto \left(\omega
+\Delta \omega\right)^2 x_1^2  + \omega^2 x_2^2$, and the new
Hamiltonian is $H+H_\epsilon$, with \be \label{Heps}
H_{\epsilon}=\frac{\epsilon}{4} \sum_m \beta_m a_{m+2}^\dagger
a_{m}^{} + (m+1)  a_{m}^\dagger   a_{m}^{} + \textrm{h.c.}, \ee
where $\beta_m=\sqrt{(m+2)(m+1)}$ and $\epsilon=\Delta \omega/
\omega$ is a small parameter.  The perturbation (\ref{Heps}) leads
to quadrupole excitations, so that states whose total angular
momenta differ by two are coupled.

\begin{figure}[h]

        \begin{center}
      \epsfig{file=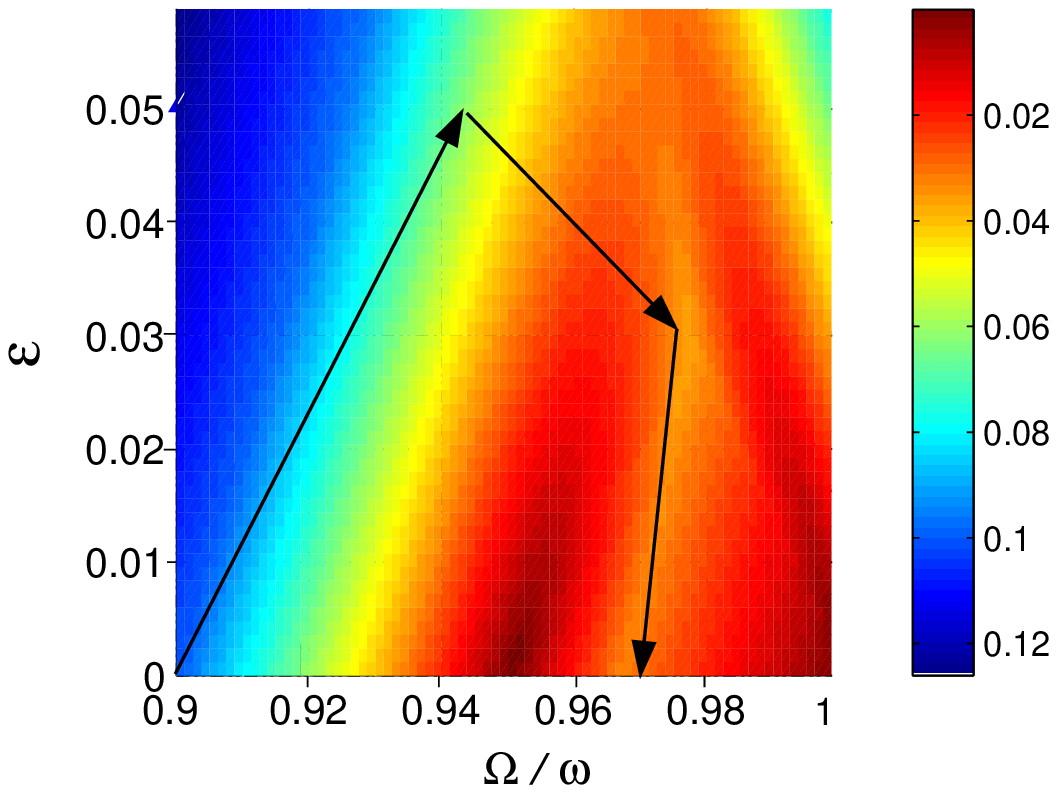,width=0.85\linewidth}
      \epsfig{file=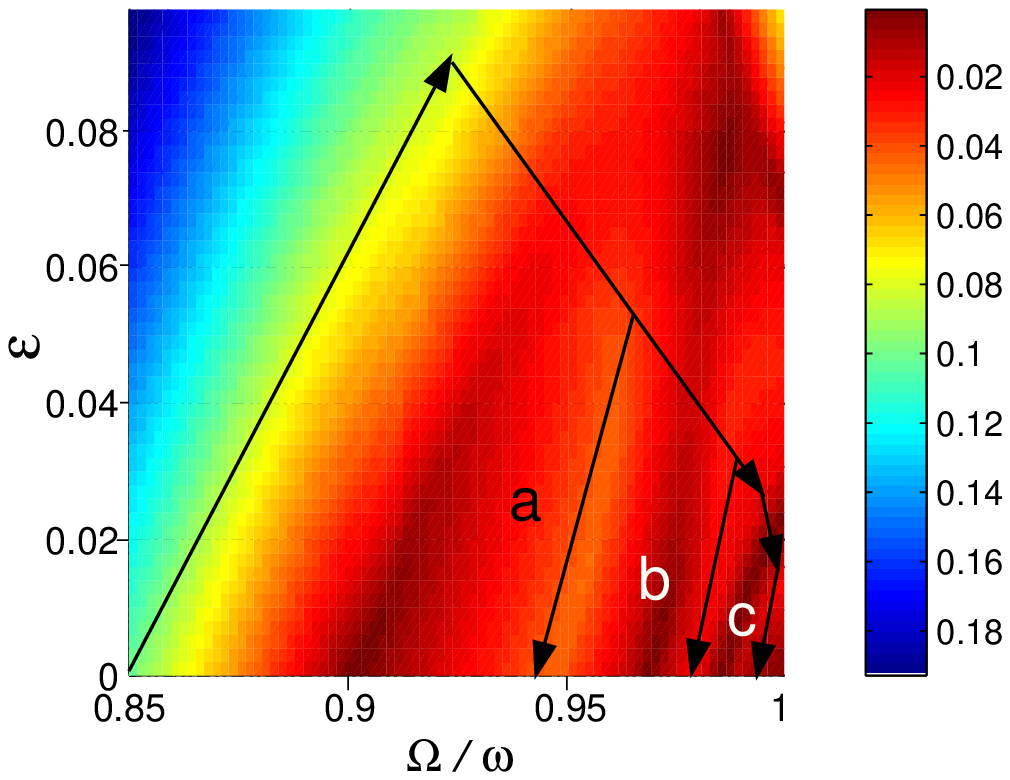,width=0.85\linewidth}

       \end{center}
\caption{Energy gap in units of $\hbar \omega$ between the ground and  first excited state
as a function of the rotation frequency $\Omega / \omega$ and the
trap deformation $\epsilon$ for an interaction strength
$\eta=0.1$. The black lines mark appropriate paths in parameter
space for adiabatic ground state evolution starting from the $L=0$ state. The adiabatic evolution times have
been calculated for a typical trapping frequency $\omega \simeq
(2 \pi) 30$kHz. Top ($N=2$): For a final fidelity
$\mathcal{F}=| \la \psi(T) | \psi_L \ra|^2=0.99$ the Laughlin
state (L=2) can be reached within T=6.5 ms. Bottom ($N=4$): Adiabatic path, evolution time T and fidelity
$\mathcal{F}$ for the following final states (see Fig.
\ref{Espec4}): (a) Pfaffian state: T=8 ms,  $\mathcal{F}=0.99$;  (b)
Quasiparticle state: T=12 ms,  $\mathcal{F}=0.99$; (c) Laughlin state:
T=215 ms,  $\mathcal{F}=0.97$.}
    \label{paths}
\end{figure}

In order to design appropriate adiabatic paths to the entangled
states described above, we have computed numerically the energy
gap between the ground and first excited state as a function of
the parameters $\Omega/ \omega$ and $\epsilon$ for $N=2$ and $N=4$ (Fig. \ref{paths}).

\begin{figure}[h]

        \begin{center}
      \epsfig{file=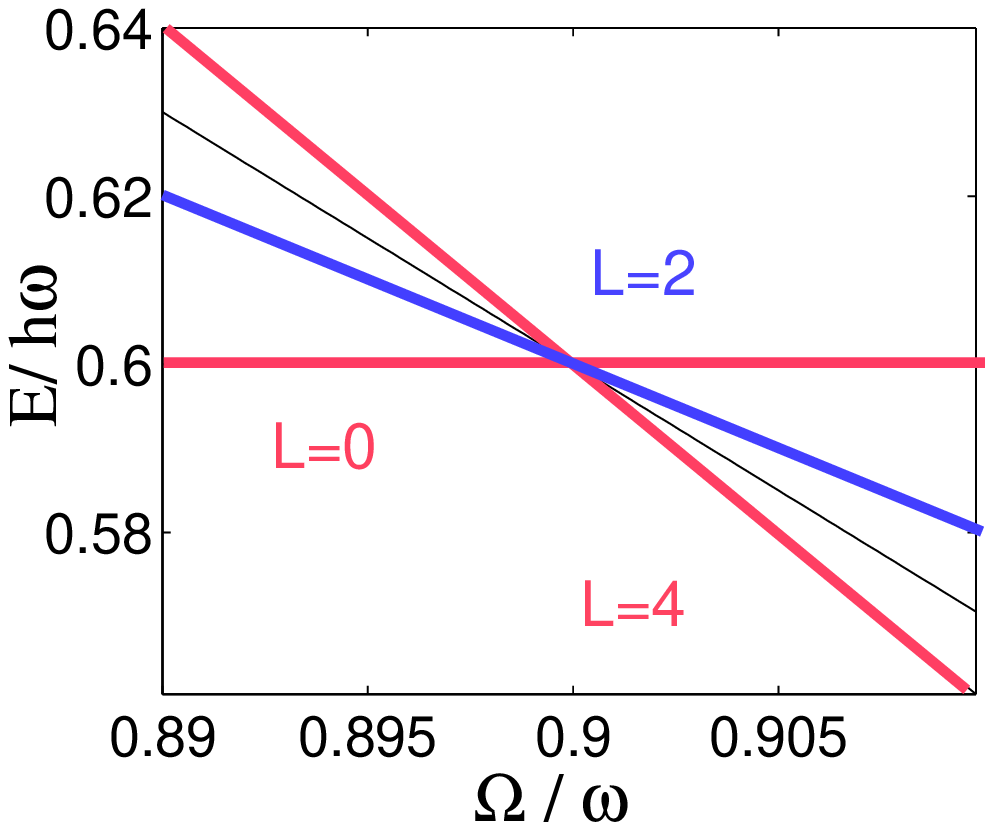,width=0.49\linewidth} 
      \epsfig{file=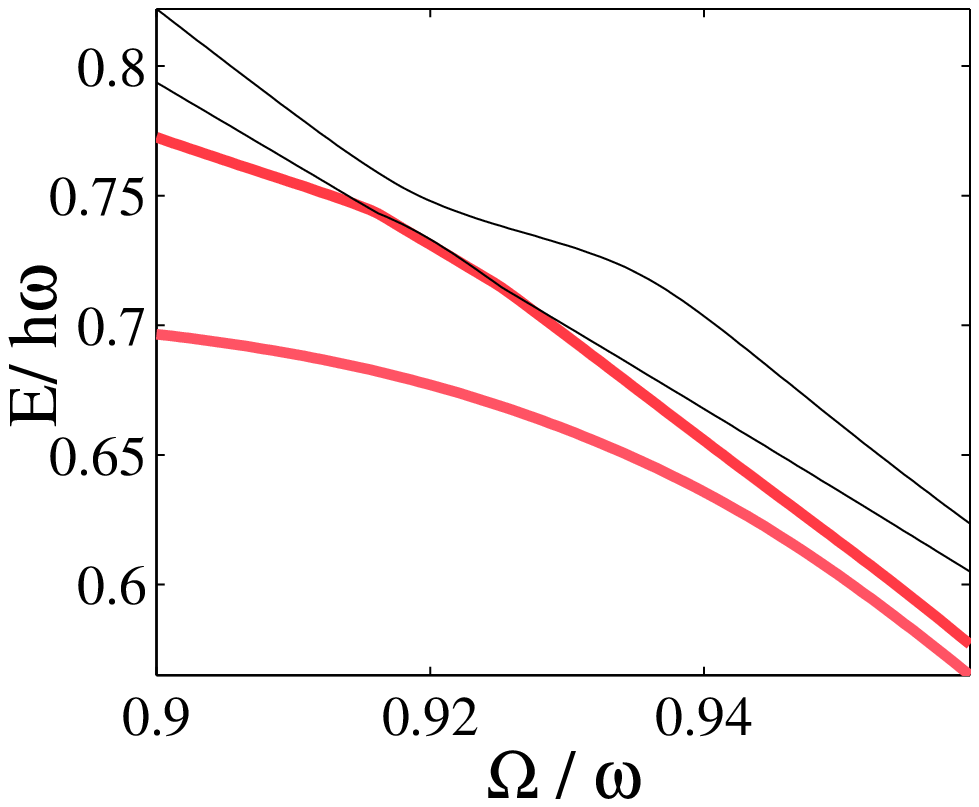,width=0.49\linewidth}
       \end{center}
\caption{Left side: Energy spectrum (in units $\hbar \omega$) for N=4 and $\eta=0.1$ in the
vicinity of the first level crossing from the L=0 to the L=4
state (see Fig. \ref{Espec4} left circle). Using quadrupole excitations ($|\Delta L|=2$) coupling
between these states is provided by the intermediate state L=2.
Right side: Emergence of an avoided level crossing for a trap
deformation $ \epsilon =0.06$.}     \label{crossing}
\end{figure}

We first note that the isolines of constant energy gap show an
approximately linear behavior. This feature can be easily
understood from a perturbative treatment of the Hamiltonian
(\ref{Heps}). To first order, the energy of states with angular
momentum $L$ is shifted by an amount $\epsilon L/4 $. Therefore
the gap profile for a given $\epsilon$ is very similar to the one
for $\epsilon=0$ but shifted an amount $\sim \epsilon$ to larger
rotation frequencies. As expected, we find that for $\epsilon \ne
0$ avoided crossings emerge (see Fig. \ref{crossing}). The energy gap of the avoided
crossings does, however, not in general increase monotonically
with the deformation $\epsilon$. Due to the interplay with other
excited states, ``saddlepoints'' appear in the gap profile, which
makes the design of appropriate adiabatic paths a nontrivial task.
For the stable entangled states of $N=2,4$ identified above these
paths are depicted in Fig. \ref{paths}. The actual time needed for
the adiabatic path depends on the number of particles as well as
on the state we want to achieve. For a typical trapping frequency
 $\omega\simeq (2 \pi) 30$kHz and an interaction coupling $\eta=0.1$,
the evolution times for the $N=2$ Laughlin state as well as for the
$L=4$ and $L=8$ states for $N=4$ are of the order of 10 ms. In
contrast, the evolution time for the $N=4$ Laughlin state is one
order of magnitude larger. We can understand this result in the
following way. For the case of $N=2$ direct coupling of the $L=0$
state to the $L=2$ Laughlin state is mediated by (\ref{Heps}). For
the case of $N=4$ there is no direct coupling between the ground
states, since their angular momenta differ by 4. But, as one can
see from the spectrum in the vicinity of the crossing to the state
$L=4$ (Fig. \ref{crossing}), there is a state with $L=2$ near
the crossing that mediates the coupling between the $L=0$ and the
$L=4$ state. A similar situation occurs for the crossing to the
$L=8$ state. However, there is no such intermediate state in
direct proximity of the crossing to the $N=4$ Laughlin state,
which leads to a decrease of the energy gap by one order of
magnitude.

Let us also comment on the situation $N=3$. Here a ground state with odd angular momentum ($L=3$) arises. From the nature of the perturbation (\ref{Heps}) it is clear that ground state evolution is not possible. However, we have shown \cite{PPCunp} that the $1/2$-Laughlin state can be reached by designing appropriate adiabatic paths via excited levels. 

\section{ Feasibility}
Let us now  discuss the  experimental
feasibility of our proposal for a small number of particles $N$. First of all we have assumed that the
lattice wells can be treated independently.  This requires that
the overlap of the Wannier functions on neighbouring sites is
small. We can estimate how intense the laser light should be in order to neglect this overlap by requiring
the size of the highest occupied angular momentum single particle state ($\approx \sqrt{2 N-1} \ell$) to be
much smaller than the separation between lattice sites, $a=\pi/k$. This leads to
the condition: $(V_0/E_R)^{1/4} \gg \sqrt{2N-1}/\pi$.   Numerical calculations of overlap integrals between adjacent sites have confirmed that for $N=2(4)$
lattice depths of $V_0/E_R \approx 30(50)$ are required. Note that these values also guarantee the validity of the harmonic approximation.
Moreover we have assumed that the available single particle states
on each well lie within the LLL. This implies that the typical
energies per particle have to be much smaller than the energy gap
to the next Landau level, $\hbar \omega$. For the limiting cases of the $L=0$ state and the Laughlin state,
this leads to the
conditions $(N-1) \eta/ 2, (N-1) (1-\Omega/ \omega) \ll 1$, which
are easily fulfilled for typical interaction strengths
($\eta\sim 0.1$) and small $N$.

Finally, in order to adiabatically achieve the entangled states identified above further conditions are required.
We analyze the most restrictive case, which corresponds to the
Laughlin state. First of
all the frequency of rotation has to be very close to the
centrifugal limit. Let us find a lower bound to the critical rotation frequency at which the
crossing to the Laughlin state appears. This can be done by calculating the rotation frequency at which the
Laughlin quasiparticle state, $\psi_{QP}([z])=\frac{\partial}{\partial z_1}
\ldots  \frac{\partial}{\partial z_N} \ \psi_L $, becomes equal in energy to the Laughlin state. Since the quasiparticle state has
$N$ units of angular momentum less than the Laughlin state and an interaction energy $\lesssim \eta$, it follows that
$\Omega_c/ \omega \geq 1- \eta/N$. For the cases of $N=2(4)$ this condition is in agreement with the exact
values found above.
Secondly, the evolution time required for the adiabatic path has to be much smaller than the typical decoherence time.
We can estimate this time in the following way. Given the critical frequency above and that the position of the avoided crossing
is displaced to larger rotation frequencies an amount proportional to $\epsilon$ it follows that the maximum $\epsilon$ we can have
is $\sim \eta/N$, corresponding to a rotation frequency $\Omega/\omega=1$.
Assuming an energy gap $\approx \epsilon$ it follows that the typical evolution time
scales as $N\eta$. For the case of $N=2(4)$ and typical $\eta$ and $\omega$ these times
are of the order of tens of miliseconds
as exactly found above, which is much smaller than the typical life time of the lattice states.
Finally, a high
degree of control of the parameters $\Omega/\omega$ and $\epsilon$
is required to perform the appropriate adiabatic paths. The
required precision scales again as $\eta/N$, which for the case of $N=4$
means a control of the parameter space up to the second digit.

 From our analysis it follows that the 
adiabatic creation of the Laughlin state 
by means of low angular
momentum excitations, as quadrupole excitations, becomes very
difficult in samples with large number of particles \cite{Boulder, Dalib04}. Even if the
centrifugal limit is possible to achieve, as it happens when
including an additional $r^4$ trapping potential \cite{Dalib04},
the adiabatic creation of the Laughlin state is still very
demanding. One reason is that the rotation frequency and the trap
deformation have to be controlled within a precision that also
scales linearly with $N$. Furthermore we point out that only the
exact knowledge of the multi-particle energy spectrum allows to
design adiabatic paths that minimize the evolution time.

\section{Detection}
We discuss now how the entangled states described
above can be detected experimentally by measuring different
properties of the states. As an important feature of our lattice
setup of independent wells, we note that any signal will be highly
enhanced by a factor equal to the number of occupied lattice sites ( $\sim 150,000$ \cite{Bloch02}).

i) \emph{Density profiles}. A very characteristic feature of the
entangled states with large angular momentum that we have
described is that they exhibit a much more extended density
distribution than the non-entangled $L=0$ state, in which the
particles are much more confined in space. For the $1/2$-Laughlin state the typical radius is given by $\bar r\approx \sqrt{2 N-1} \ \ell$. In the case of 
$N=2(4)$ this results in a  radius that is $\sim 2(3)$ times
larger than in the case of the condensate. As proposed in
\cite{RC03} the density profile of states within the LLL can be
measured in a time of flight (TOF) image of the atomic system,
since the momentum distribution coincides with the density profile
for LLL states. In our case  of independent
3D wells, a TOF absorption picture after expansion time $t$  will exhibit a broad central peak of the form:
 \be \label{rhotf}
 \rho({\bf r},
t)\approx\frac{N_{s}}{(\omega t)^3 } \ |\rho_0( -i z / (
\omega t), x_3/ (\omega t)|^2 \ . 
\ee 
Here,  $\rho_0(z,x_3)$ is the initial density distribution on a single well. In the TOF image it is enhanced by a factor proportional to the number of lattice sites $N_s$ and rescaled by a factor $\omega t \gg 1$. The $\pi/2$ rotation $z\rightarrow -i z$ leaves isotropic states, like the FQH states described above, unaffected.
The underlying assumption of free
(interactionless) expansion is justified, since the interaction
energy is small compared to the kinetic energy (in the stationary
frame).

ii)\emph{Angular momentum}. For any state within the
LLL integration over the density distribution gives $\int d{\bf
r}\ r^2 \rho({\bf r})=L+N$. Thus in the limit of weak interaction the total  angular momentum can be extracted directly from the TOF picture.

iii)\emph{Correlation functions}. Here we propose a novel technique that
makes directly use of the rich possibilities offered by the
optical lattice setup and which allows to measure both the
$g_1=\langle \psi^\dagger ({\bf r})  \psi ({\bf r'})\rangle$ and
$g_2=\langle \psi^\dagger ({\bf r})  \psi^\dagger ({\bf r'})
\psi({\bf r})\psi ({\bf r'}) \rangle$  correlation functions. The
$g_2$ correlation function is for instance very characteristic for
a Laughlin state. Since particles can only be at least in relative
angular momentum $m_r=2$ it follows that $g_2\propto \lvert
r-r'\vert^4$. This behavior  reveals  the
$\frac{1}{2}$-fractional nature of this Laughlin state. 

We consider two species $a$ and $b$ (hyperfine levels) of bosonic
atoms, which can be coupled via Raman transitions. We start with
atoms in level $a$ and create the entangled state of interest $|
\Psi_{i} \rangle$ with the method described above. Next we apply a
$\pi/2$-pulse with the laser and create an equal superposition of
$a$ and $b$ states. Finally, we shift the lattice potential trapping atoms
of type $b$ (as proposed in \cite{Jaksch00} and realized in
\cite{Bloch03}) by a distance ${\bf r_0}$ small compared to the
lattice spacing  and perform another $\pi/2$-pulse. In the Heisenberg picture this procedure corresponds to the following transformation of the field operator for  species $a$: 
\be
\psi_a({\bf r}) \rightarrow \psi_a({\bf
 r})+\psi_a({\bf r+r_0}) \ .
\ee 
  Thus the density distribution of 
atoms of type $a$ in this new state $| \Psi_{f} \rangle$ contains information
about the $g_1$ correlation function of the original state:
\bea
& &\langle \Psi_f|\psi_a^\dagger({\bf r})\psi_a({\bf
r})|\Psi_f\rangle= \\
& &\langle \Psi_i|(\psi_a^\dagger({\bf r})
+\psi_a^\dagger({\bf r+r_0})) (\psi_a({\bf r})+ \psi_a({\bf
r+r_0}) )|\Psi_i\rangle \nonumber  \ .
\eea
 Using this procedure we can also measure
higher order correlation functions like $g_2$. In this case
measuring the interaction energy of the final state will allow us to
calculate the $g_2$ of the initial state. For instance, for the
Laughlin state we have:
\bea
& &  E_{int}({\bf r_0})= \\
& &\frac{\pi \eta}{4} \int d{\bf r} \langle \Psi_i |
 \psi_a^\dagger({\bf r})\psi_a^\dagger({\bf r+r_0})\psi_a({\bf
 r})\psi_a({\bf r+r_0}) | \Psi_i \rangle \ .  \nonumber
\eea 
The interaction energy is, unfortunately, not directly accessible
experimentally. However, the total energy of the final state can
be obtained from integrating over the TOF absorption picture,
since energy is conserved during the time of flight. For small
coupling $\eta$, however, the measurable effect due to
interactions will be small compared to the kinetic part of the
energy. In addition, the kinetic energy itself shows a significant
dependence on the shifting $r_0$, which has to be distinguished
from the interaction. Hence, we propose to tune the scattering
length $a_s$ (e.g. via a photo association induced Feshbach
resonance \cite{Ketterle99}) and to measure the interaction energy
both in the weak and strong scattering limit. The difference would
then reveal the characteristic behavior of the $g_2$ correlation
function.

We finally note that, as a further way of detection for the $N=4$
Laughlin state, a strong reduction of the three body losses should
be observed.
\section{conclusion}
In conclusion, we have shown how to motionally entangle a small
number of particles into a sequence of interesting FQH states. We
have fully addressed the adiabatic creation of these states and
proposed new techniques for their experimental detection.
\section{Acknowledgements}
We acknowledge helpful discussions with I. Bloch, J. Garc\'ia-Ripoll and M. Greiner. This work was supported in part by EU IST projects (RESQ and QUPRODIS), the DFG, and the Kompetenznetzwerk ``Quanteninformationsverarbeitung'' der Bayerischen Staatsregierung.


\begin{thebibliography}{10}
\bibitem{Com} Typically the filling factor $N$ is not homogeneous,
because the Gaussian intensity profile of the laser beams results
in an additional weak harmonic confinement of the atoms. However, it has
been demonstrated that commensurate  filling can be achieved using an  entanglement interferometer \cite{Bloch04}.
\end{thebibliography}
\end{document}